\documentclass[a4paper,8pt]{article}
\usepackage[cp1251]{inputenc}
\usepackage[russian,english]{babel}
\usepackage{array,longtable}
\usepackage{amssymb, latexsym, amsthm, amsmath, amsxtra}
\usepackage{graphicx}

\graphicspath{{}}
\DeclareGraphicsExtensions{.pdf,.png,.jpg,}

\makeatletter

\begin{document}

\large

\begin{center}
\title{}{\bf  On the soliton model of azimuthal - anisotropic elliptical flows of quark-gluon plasma and hadron jets.  }
\vskip 1cm

\author{}
 R.K. Salimov \textsuperscript{1}, T.R. Salimov \textsuperscript{2}
{}

\vskip 1cm
{ \textsuperscript{2} Bashkir State University, Ufa, Russia }

{ \textsuperscript{1} Moscow Institute of Physics and Tehnology, Dolgoprudny,  Russia}


\vskip 0.5cm
e-mail: salimovrk@bashedu.ru

\end{center}

\vskip 1cm

{\bf Abstract}
 \par
The paper presents a system of nonlinear Lorentz-invariant equations. The behavior of their localized solutions is analogous to the processes of formation of azimuthal-anisotropic elliptical flows of quark-gluon plasma and hadron jets. In contrast to the hydrodynamic models of an ideal fluid, the elliptic flow in this model has a significant negative acceleration. The presence of the negative acceleration allows us to propose a hypothesis about the direct elliptic flow of photons as braking radiation QGP.
 \par
 \vskip 0.5cm


\par
\vskip 1cm

\begin{center}
{\bf Introduction}
\end{center}

In the phenomenological description of azimuthal-anisotropic elliptical flows arising in collisions of nuclei, the resulting quark-gluon plasma is often considered as a kind of liquid. The hydrodynamic description of the formation and further evolution of quark-gluon plasma (QGP) in the high-energy region is quite successful [1-10]. To simulate elliptical flows in this case, relativistic hydrodynamics is used [11-15]. However, in recent years it has become clear that the hydrodynamic description does not fully describe the experimental data [16,17]. In particular, the elliptic flow v2 of direct photons contradicted theoretical expectations [ 18-20].

Recent studies of Lorentz-invariant 2D nonlinear equations [21] with fractional nonlinearity show that their localized solutions have hydrodynamic properties, in particular, surface tension. This manifests itself in fluctuations in the form of solutions. Another interesting property of long-lived localized solutions of such equations is their nontrivial form and nonuniform distribution of the field amplitude. This nonuniform distribution can have several local maxima. Such properties make it possible to consider such solutions as some kind of soliton models of hadrons. We also note that similar high-energy solutions break down into several localized solutions of lower energy. In addition, angular anisotropy is observed during this decay.

For physics applications, 3D+1 equations are more interesting. Therefore, this article considers the numerical solution of 3D+1 nonlinear Lorentz-invariant equations with fractional nonlinearity. The purpose of this work is to test the assumption about the azimuthal anisotropy of the localized solution of such equations under the initial conditions in the form of an ellipse and clarifying the differences between this model and the hydrodynamic one.

\begin{center}
{\bf Results}
\end{center}

A system of equations of the form:

\begin{align}
 u_{xx}+u_{yy}+u_{zz}-u_{tt}=\alpha(u+u(u^2+v^2))+\beta\frac{u}{(u^2+v^2)^\frac{n}{2n+1}}+\gamma\frac{u}{(u^2+v^2)^\frac{m}{2m+1}}
 \nonumber\\-\zeta sin(u)cos^2(v)+\eta v \label{eq:3}
   \end{align}

\begin{align}
 v_{xx}+v_{yy}+v_{zz}-v_{tt}=\alpha(v+v(u^2+v^2))+\beta\frac{v}{(u^2+v^2)^\frac{n}{2n+1}}+\gamma\frac{v}{(u^2+v^2)^\frac{m}{2m+1}}
 \nonumber\\-\zeta sin(v)cos^2(u)+\eta u \label{eq:4}
   \end{align}
was studied.

Let us now explain the need to use such a number of summands. Two scalar fields $u, v$ are considered since numerical simulation in the two-dimensional case shows that without the condition of cylindrical symmetry localized solutions from one field u are unstable and decay rather quickly ($t\backsim 30$). The decay of localized solutions  occurred when the amplitude of the solution passed through zero values. Therefore, to obtain stable localized solutions, equations for two scalar fields were considered. Further, the summand at the coefficient $\alpha$  is quite significant and ensures the impossibility of a long existence of localized solutions of large amplitude.  Localization of solutions at small amplitudes is ensured by the summands at the coefficients $\beta$ and $\gamma$. Indeed, for nonzero fields u and v, the summands at the coefficients $\beta$ and $\gamma$ are potential barriers of the form  $\frac{\beta}{(u^2+v^2)^\frac{n}{2n+1}}$. Numerical experiments show that the presence of two such summands with different degrees increases the localization of solutions in comparison with the case of one such summand. The summand at the coefficient $\zeta$ also ensures the localization of solutions. The summand at the coefficient $\eta$ leads to the fact that for a nonzero field $u$ a field $v$ necessarily appears, which also contributes to the existence of localized long-lived solutions.

In the numerical study, a modified Christiansen – Lomdal difference method [22] of fourth-order accuracy was used. Numerical modeling was performed in a 15x15x15 cube. The step along the coordinate h = 0.06.
At first, the system of equations (1) (2) was considered with the initial conditions in the form of an ellipse,
\begin{align}
u(x,y,z,0)=2.4\exp(-2((x'-x'_0)^2/a^2+(y'-y'_0)^2/b^2+(z'-z'_0)^2/c^2))
 \end{align}

\begin{align}
v(x,y,z,0)=0
 \end{align}

compressed in two directions, for example: $a = 0.1, b = 0.11, c = 1.3$.

 The simulation results show that the solution spatially expands over time in the x'y 'plane, while breaking into “drops” (fig.1).

\begin{figure}[h]
\center
\includegraphics[width=11cm, height=10cm]{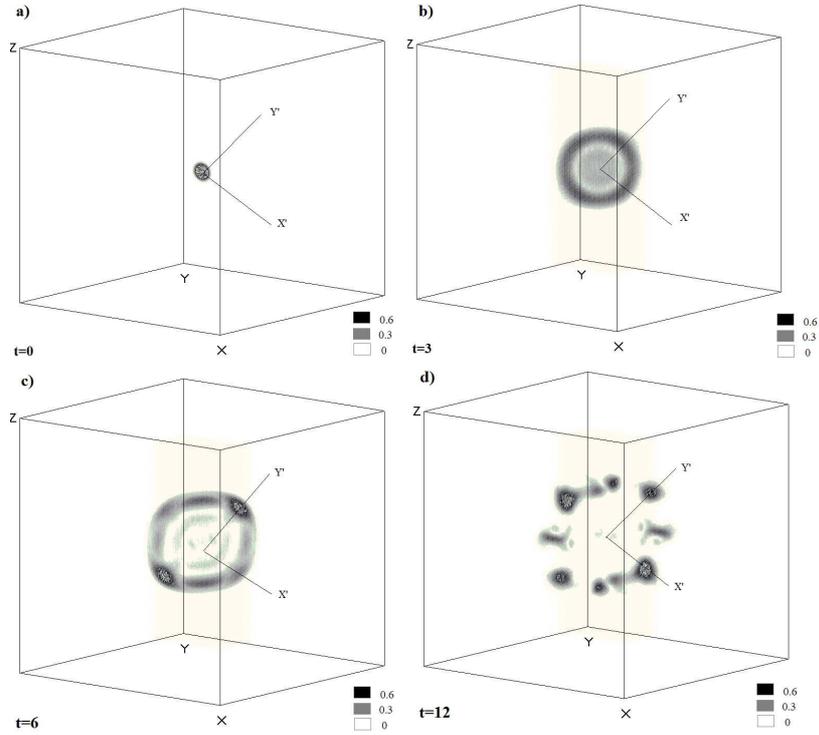}
\caption{Spatial distribution of the value $(u^2+v^2)^{1/2}$ for parameters. $n=4;m=2;\alpha=12;\beta=1,7;\gamma=1.8;\zeta=1.2;\eta=4$. }
\label{schema}
\end{figure}

Such solutions can be regarded as a phenomenological model of azimuthal-anisotropic elliptical flows arising from nuclear collisions. Let us now consider in more detail the dynamics of an elliptic flow. Figures 1a, 1b show that at the initial moment of time ($t = 0 \backsim 3$) the periphery of the elliptic flow moves at a high speed $v = 0.74$ see Fig. 1a, 1b. At subsequent moments of time, the movement slows down noticeably, $v = 0.32$ see Fig. 1b, 1c. After the process of fragmentation and localization of the solution, the speed of the fragmented “drops” does not change see Fig. 1d. The same behavior of solutions is observed for the parameters close to the parameters of the solution presented. That is, in this model, there is a negative acceleration of the elliptic flow.

\begin{center}
{\bf Discussion}
\end{center}

From the results presented it can be seen that the proposed model qualitatively describes the process of fragmentation and localization of a quark - gluon “liquid” into “drops”. That is, it qualitatively describes the process of formation of azimuthal-anisotropic flows of quark-gluon plasma.  An analogy with hydrodynamics is provided by the summands at the coefficients $\alpha,\beta ,\gamma$ in equations (1-2). The summand at the coefficient $\alpha$ creates a certain excess “pressure” at large amplitudes of the solution. And the summands at the coefficients $\beta, \gamma$ provide the localization of solutions, being an analogue of the “surface tension”.

A specific feature of this soliton model is the presence of negative acceleration of the elliptic flow. A similar feature of this model allows us to propose a hypothesis about the direct elliptic flow of photons as braking radiation QGP. Thus, this model offers the prospect of a qualitative explanation of the puzzle of the direct elliptic photon flow. According to the authors, the presented model is also interesting from a methodological point of view and is worthy of further study.


\end{document}